\documentclass[aps,prd,preprintnumbers, nofootinbib]{revtex4}
\usepackage{bm}
\usepackage{latexsym}
\usepackage{amsmath,amsfonts,amssymb}
\usepackage{graphicx,epsfig}
\usepackage{psfrag}
\usepackage{amsthm}
\usepackage{color}

\usepackage{amsmath}
\usepackage{braket}
\usepackage{indent first}
\usepackage[normalem]{ulem}
\useunder{\uline}{\ul}{}

\begin{document}
\title{Information recovery from evaporating rotating charged black holes}

\author{Zhi-Wei Wang}
\email{zhiweiwang.phy@gmail.com}
\affiliation{Department of Physics and Astronomy, University of Lethbridge, 4401 University Drive, Lethbridge, Alberta T1K 3M4, Canada}

\author{Samuel L. Braunstein}
\email{sam.braunstein@york.ac.uk}
\affiliation{Computer Science, University of York, York Y010 5GH, United Kingdom}

\author{Saurya Das\footnote{Author for correspondence (saurya.das@uleth.ca)}}
\email{saurya.das@uleth.ca}
\affiliation{Theoretical Physics Group and Quantum Alberta, Department of Physics and Astronomy, University of Lethbridge, 4401 University Drive, Lethbridge, Alberta T1K 3M4, Canada}

\begin{abstract}
In classical gravity, nothing can escape from a 
black hole, not even light.
In particular, this happens for
stationary black holes
because their horizons are null. 
We show, on the other hand, 
that the apparent horizon and the region near $r=0$ of an 
{\it evaporating} charged, rotating black hole are both timelike. 
This implies that there exists a channel, via which classical or quantum information can escape to the outside, as the black hole evaporates. 
Since astrophysical black holes have at least some rotation, our results apply to  all black holes in nature. We discuss implications of our result. 
\end{abstract}

\maketitle

\section{Introduction}

In classical general relativity, a black hole is a region of spacetime from which  nothing escapes. This result 
underwent a change with 
the discovery of Hawking radiation \cite{Hawking1974,Hawking1975}.  
Studying the evolution of a vacuum quantum state in curved spacetime with a horizon, Hawking showed that the ingoing
vacuum state has a probability to yield a non-vacuum outgoing state when the former 
enters the horizon of a black hole.
%
This means that a black hole can radiate energy to infinity via this quantum mechanism.

Now, both classical and quantum information can enter the horizon of a black hole. 
However, as Hawking's work showed,
only thermal Hawking radiation carrying very little information emerges from it \cite{Hawking1976}. 
Therefore, when the black hole evaporates completely, all information carried by the matter that was used to create the black hole is apparently lost forever. 
This violates unitarity in quantum theory which requires the information of a quantum system be conserved during its evolution. Further study of this apparent information loss paradox led to the idea of black hole complementarity \cite{Hooft1985,Hooft1990,Susskind1993}, the firewall paradox \cite{Braunstein2009,Almheiri2013} etc. 

When discussing the above problem, 
people often overlook an implicit hypothesis that the causal structure of a black hole is not disturbed by Hawking radiation itself. 
Parikh and Wilczek were the first to 
examine this assumption and show that the singularity of an evaporating charged black [white] hole may be accessible for an observer outside when the metric is time-dependent \cite{Parikh1999}. However, the metric used in their paper in fact seems to represent a white hole, contrary to their claim that it is a black hole (see Appendix A for a detailed analysis). Recently, numerical simulation also shows that the horizon of a charged black hole may be timelike \cite{Schindler2020}. 
We know on the other hand, that much like other astronomical objects, all black holes in nature are rotating and uncharged, and the probability of finding a black hole with zero rotation is practically nil. 

Motivated by the above, 
in this work \cite{Braunstein2021}, we show that for a rotating and charged black hole which is Hawking radiating, there is a classical channel through which information can escape, and following the above reasoning, it provides an extended window of information recovery from its interior. In the process, the black hole shrinks of course, but presumably at a faster rate than 
predicted by Hawking radiation, because of the additional outflow of information and associated matter. 

We demonstrate the above rigorously by constructing the Penrose diagram for the above 
process. In particular, we prove that the region immediately surrounding $r=0$ is timelike, and the apparent horizon is timelike as well. 
These two results and the Penrose diagram all together imply null geodesics originating from anywhere near the centre of the black hole 
to the apparent horizon may emerge to the outside Universe. This in turn provides a route for classical or quantum information to escape from the black hole. This escape of potentially a large amount of information 
must be taken into account in any attempt to resolve the information loss problem. What is most significant is perhaps the fact that the escaping information need not be thermal.  

\section{Coordinates for an evaporating rotating charged black hole}

Since we are interested in evaporating black holes, we will start with the following time-dependent metric, which represents a rotating and charged Vaidya-type
black hole \cite{Vaidya,jing}, as shown in Appendix B
\begin{eqnarray}
\!ds^2\!&=&-\biggl(1-\frac{2Mr-Q^2}{\sigma^2}\biggr)du^2 + 2du\,
dr + \sigma^{2}d\theta^2 \label{rotmetric} - 2a\sin^2\theta\, dr\, d\phi \nonumber \\
&&-\frac{2(2Mr-Q^2) a }{\sigma^2}\sin^2\theta\, du\, d\phi
+\biggl(\frac{2Mr-Q^2}{\sigma^2}\,a^2\sin^2\theta
+r^2 + a^2\biggr)\sin^2\theta\, d\phi^2, \nonumber \\
\end{eqnarray}
where $\sigma^{2}\equiv r^2+a^2\cos^2\theta$, and $M=M(u)$, $Q=Q(u)$
and $a=a(u)$ denote smooth decreasing functions of the retarded time $u$.

Next, to construct a two-dimensional Penrose diagram for the above metric, we restrict ourselves to the symmetry axis
along $\theta=0$ \cite{Carter66}. 
Then one gets, from Eq.~(\ref{rotmetric}):
\begin{eqnarray}
ds^2 = -\biggl( 1 - \frac{2Mr - Q^2}{r^2 + a^2}\biggr) du^2
{+} 2\, du\, dr = -\biggl[ \frac{r^2-2Mr+a^2+Q^2}{r^2+a^2}\, du {-} 2\, dr\biggr] du.
\label{metric2}
\end{eqnarray}
We also assume without loss of generality that
$M$, $a$ and $Q$ are proportional to each other and linear functions of $u$
%
\begin{eqnarray}
M(u) &=& M_0 ~(\mbox{constant}),\quad \; ~ u<u_0 \label{mass0} \\
&=& \mu u +b  
\equiv \tilde u,   
\qquad\quad u_0< u \leq -b/\mu, \mu <0  \label{mass1} \\
&=& 0,\qquad\qquad\qquad~~ -b/\mu < u\\
a(u) &=& \lambda_1 M(u), \qquad\quad~~~~ 0 \leq \lambda_1 \leq 1\\
Q(u) &=& \lambda_2 M(u), \qquad\quad~~~~ 0 \leq \lambda_2 \leq 1 . \label{mass2}
\end{eqnarray}
This means that the black hole starts its process of evaporation at time $u=u_0$, and undergoes a mass decrease at a constant rate $\mu$.
Contrary to Ref.~[\citenum{jing}], where the 
angular momentum was held fixed, in this work we make it time-dependent.
This is because on the one hand, 
the form of the metric continues to hold with that generalization and moreover,
the condition $\lambda_1, \lambda_2 \leq 1$ ensures that there are no naked singularities.

The following points may be noted here. 
First, continuity of the functions $M(u),a(u),Q(u)$ in Eqs.~(\ref{mass0}-\ref{mass2}) guarantees that the various patches of the 
Penrose diagram that will smoothly join to each other \cite{Parikh1999}. 
Second, considering arbitrary decay rates (i.e. not constant over time) is also straightforward.
In this case, one can simply break down the non-linear decay function such as 
$M(u)$ into a series of linear functions of the form
$M(u) = \sum_{i=1}^N (\mu_i u + b_i) 
\left[ \Theta( u_i+ \frac{1}{2}) - \Theta( u_i+ \frac{1}{2}) \right]$
($u_0 \leq u_i 
$), and similarly for
$a(u)$ and $Q(u)$, for $N\gg 1 $. 

\begin{figure}[htp] 
\centering
{\includegraphics[width=0.35\textwidth]
{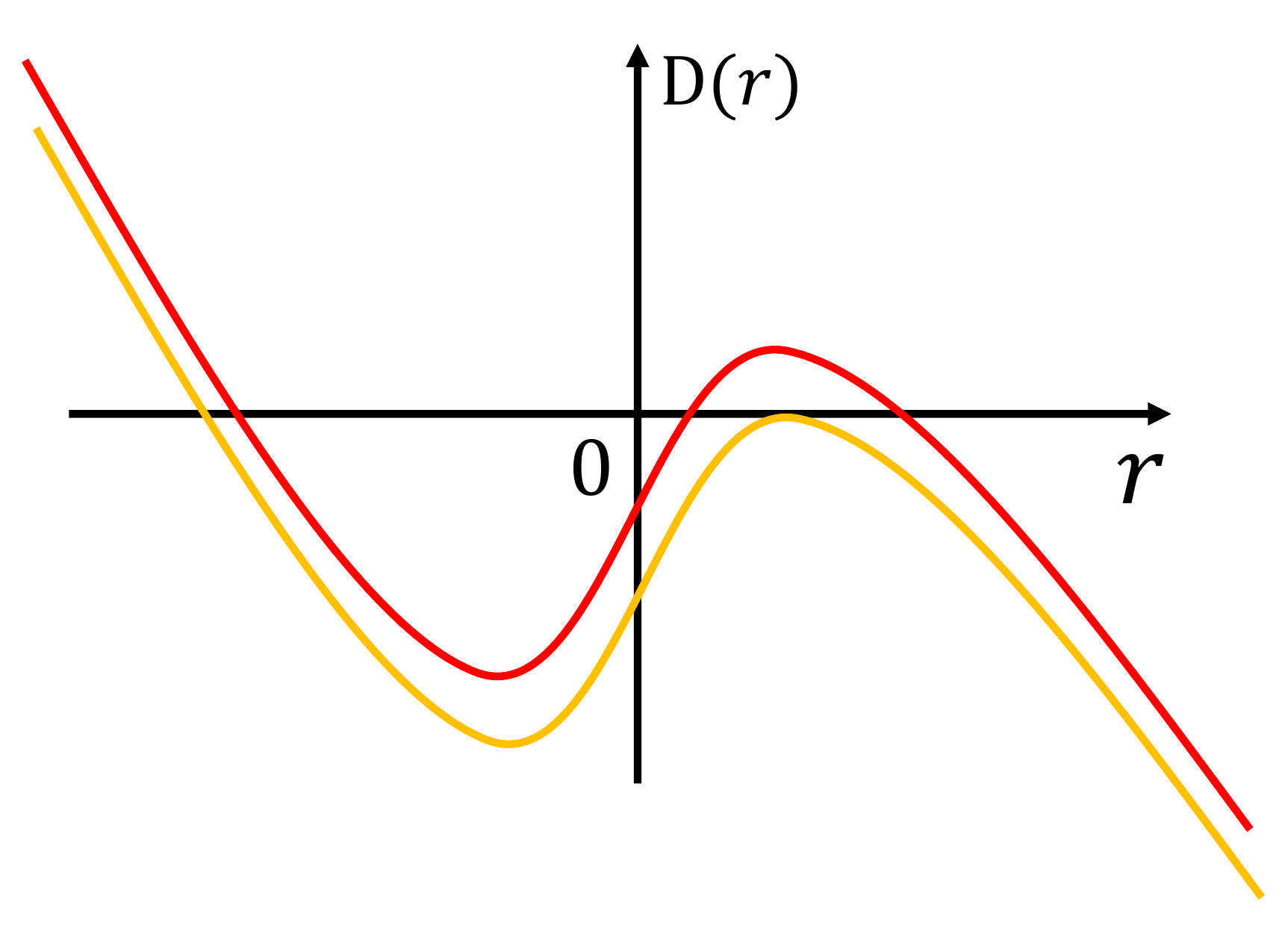}}
\vskip -0.1truein
\caption{Scenarios caused singularities for Eqs.~(\ref{v1}) and (\ref{v2}).}
\label{cubic}
\end{figure}

Now, following Ref.~[\onlinecite{Parikh1999}] and 
details presented in Ref.~[\onlinecite{Braunstein2021}], 
we find a set of coordinates which are smooth across the horizon. To this end, 
we first write metric (\ref{metric2}) in the so-called `double null' form:
\begin{eqnarray}
ds^2 = - \frac{g(\tilde u,r)}{\mu} d\tilde u dv ~,
\end{eqnarray}
where, 
\begin{eqnarray}
dv \equiv \frac{1}{g(\tilde u, r)} \left[ \left( \frac{r^2 - 2Mr +
(\lambda_1^2 + \lambda_2^2)M^2}{r^2 + {\lambda_1}^2 M^2}  \right) \frac{d\tilde u}{{\mu}} {-} 2dr \right].
\label{metric3}
\end{eqnarray}
with the integrating factor
\begin{eqnarray}
g(\tilde u,r) =  \left( \frac{r^2 - 2Mr +
(\lambda_1^2 + \lambda_2^2)M^2}{r^2 + {\lambda_1}^2 M^2}  \right) \frac{\tilde u}{{\mu}} {-} 2r~.
\end{eqnarray}
We may then read off  
\begin{eqnarray}
\frac{\partial v}{\partial r} = \frac{r^2 + \lambda_1^2 \tilde u^2}{{-}r[r^2 + \lambda_1^2 \tilde u^2]
+ \frac{\tilde u}{2{\mu}} \left[ r^2 - 2\tilde u r + (\lambda_1^2 + \lambda_2^2)\tilde u^2 \right] } \label{v1} \\
\frac{\partial v}{\partial \tilde u} =
\frac{ \frac{1}{2a} ( r^2 - 2Mr + (\lambda_1^2 + \lambda_2^2)\tilde u^2) }{{-}r[r^2 + \lambda_1^2 \tilde u^2]
+ \frac{\tilde u}{2{\mu}} \left[ r^2 - 2\tilde u r + (\lambda_1^2 + \lambda_2^2)\tilde u^2 \right] } ~. \label{v2}
\end{eqnarray}
As one can see, the 
common denominators $D(r)$ of the above expressions are cubic in $r$, which may be rewritten as
\begin{eqnarray}
D(r)&=&-r[r^2 + \lambda_1^2 \tilde u^2]
+ \frac{\tilde u}{2{\mu}} \left[ r^2 - 2\tilde u r + (\lambda_1^2 + \lambda_2^2)\tilde u^2 \right]  \nonumber \\
&=& -r^3 - \lambda_1^2 \tilde u^2 r
+ \frac{\tilde u}{2{\mu}} r^2 - \frac{\tilde u^2}{\mu} r + \frac{\tilde u^3}{2{\mu}} (\lambda_1^2 + \lambda_2^2)  \nonumber \\
&=& -r^3 + \frac{\tilde u}{2{\mu}} r^2 - (\lambda_1^2 \tilde u^2 +\frac{\tilde u^2}{\mu}) r + \frac{\tilde u^3}{2{\mu}} (\lambda_1^2 + \lambda_2^2).   \nonumber \\
\end{eqnarray}
Since $D(-\infty)>0$, $D(+\infty)<0$, and $D(0)= \frac{\tilde u^3}{2{\mu}} (\lambda_1^2 + \lambda_2^2) < 0$ because of $\mu<0$ and $\tilde u > 0$ ($\tilde u = 0$ represents a flat spacetime and hence is ignored here), the scenarios with positive roots for the cubic equation $D(r)=0$ are well described by Fig.~\ref{cubic}. 
No other scenario has a positive root. 

For the case represented by the red line in Fig.~\ref{cubic}, the denominators have three roots $r_1,r_2,r_3$,
(with $r_1 >r_2 >0>r_3$) where $v$ has coordinate singularities, and 
\begin{eqnarray}
v &=& \sum_{i=1}^{3} A_i \ln(r-r_i)~,
\label{v4}
\end{eqnarray}
with
\begin{eqnarray}
A_1 &=& \frac{r_1^2}{(r_1-r^2)(r_1-r_3)} + \lambda_1 \tilde u^2 , \qquad
A_2 = \frac{-r_2^2}{(r_1-r^2)(r_2-r_3)} + \lambda_1 \tilde u^2, \nonumber \\ 
A_3 &=& \frac{r_3^2}{(r_1-r_3)(r_2-r_3)} + \lambda_1 \tilde u^2 ~.
\end{eqnarray}
Then the complete set of singularity free coordinates for the two patches may be defined as:
\begin{eqnarray}
V_2(v) &\equiv & e^{v/A_1} \nonumber \\ 
&=& (r-r_1) (r-r_2)^{A_2/A_1}(r-r_3)^{A_3/A_1}~,\qquad\qquad\;\;\; r_2 \leq r < \infty  \\
V_1(v) &\equiv & k_1 - (-V_2)^{A_1/A_2} \nonumber \\ 
&=& k_2 + (r_1-r)^{A_1/A_2} (r_2-r) (r-r_3)^{A_3/A_2}~,\qquad\quad 0 \leq r < r_2
\label{v5}
\end{eqnarray}
The constants $k_i$ are determined by matching $V_2$ with $V_1$ in
$r_2 < r < r_1$.
If the denominator $D(r)$ of Eq.~(\ref{v1}) is 
described by the gold line in Fig.~\ref{cubic}, it has a negative root $r_3$ and a positive double root $r_1$ such that it can be
factored as $(r-r_3)(r-r_1)^2$. This is nothing but a special case that $r_1=r_2$ in the red-line scenario in Fig.~\ref{cubic}. 

\section{Timelike regions}

With the results in last section, it is easy to show that 
$r=0$ and its immediate neighbourhood is 
{\it timelike}. 
For example, in the case of Eqs.~(\ref{v4}-\ref{v5}) above,
\begin{eqnarray}
ds^2 &=& - \frac{g(\tilde u,r) A}{V(\tilde u,r) {\mu} }~d\tilde u\, dV  \\
&\stackrel{\smash{r\rightarrow 0}}{\longrightarrow}& -\left( \frac{\lambda_1^2 + \lambda_2^2}{\lambda_1^2}\right) du^2 <0 ~.
\label{ds2}
\end{eqnarray}
It may be noted that the `ring singularity' of the metric is
at $r=0$ {\it and} $\theta=\pi/2$. It follows that except for that singular point, the region in the neighbourhood of $r=0$ is regular and of finite curvature, except perhaps at the end of the evaporation process.

We now consider the apparent horizon and show that it is 
timelike as well. We follow the 
procedure of Ref.~[\onlinecite{kami}], 
and define the function:
\begin{eqnarray}
f(u,r)= r - R_\pm ~,
\label{funct}
\end{eqnarray}
where $R_\pm$ represents the outer and inner apparent horizons of the evaporating rotating charged black hole.
A rigorous calculations for $R_\pm$ yields
\begin{equation}
R_\pm = r_\pm + \frac{a \lambda_1 (r_\pm^2+a^2)}{r_\pm^2-(M+2a\mu \lambda_1)r_\pm} \mu + O(\mu^2) ,
\label{evaphorManu}
\end{equation} 
where 
\begin{eqnarray}
r_\pm = M \pm (M^2 - Q^2 - a^2)^{1/2}
\end{eqnarray}
(see Appendix B for more details).

We assume that the evaporating process is sufficiently slow and that $|\mu|$ is sufficiently small. Therefore, we focus on the leading term in Eq.~(\ref{evaphorManu}) for our future calculations. Thus, Eq.~(\ref{funct}) reduces to
\begin{eqnarray}
f(u,r)= r - M \mp (M^2 - Q^2 - a^2)^{1/2} ~,
\end{eqnarray}
such that the apparent horizons are at
\begin{eqnarray}
f(u,r)=0, ~r_\pm = M \pm (M^2 - Q^2 - a^2)^{1/2}~.
\end{eqnarray}

Now since $M$, $Q$ and $a$ are functions of $u$ 
the vector $n_a$, normal to the surface $f(u,r)=0$, will have components given by
\begin{eqnarray}
n_u = f_{,u} 
= {-} \mu {\mp} \frac{\mu~( M - \lambda_1 a - \lambda_2 Q )}{ (M^2 - Q^2 - a^2)^{1/2} } ~,\qquad
n_r = f_{,r}=1~,
\end{eqnarray}
and norm
\begin{eqnarray}
n^2 = g^{ab}n_a n_b = 2 { \mu [-1 \mp (1 -\lambda_1^2 -\lambda_2^2)^{1/2} } ] >0~,
\label{n2}
\end{eqnarray}
where we have used $\mu<0$ from Eq.~(\ref{mass1}).
Therefore, $n_a$ is spacelike and the apparent horizon is timelike.

\section{Penrose diagram}

\begin{figure}[htp]
\centering
\vskip -0.1truein
\includegraphics[width=0.35\textwidth]{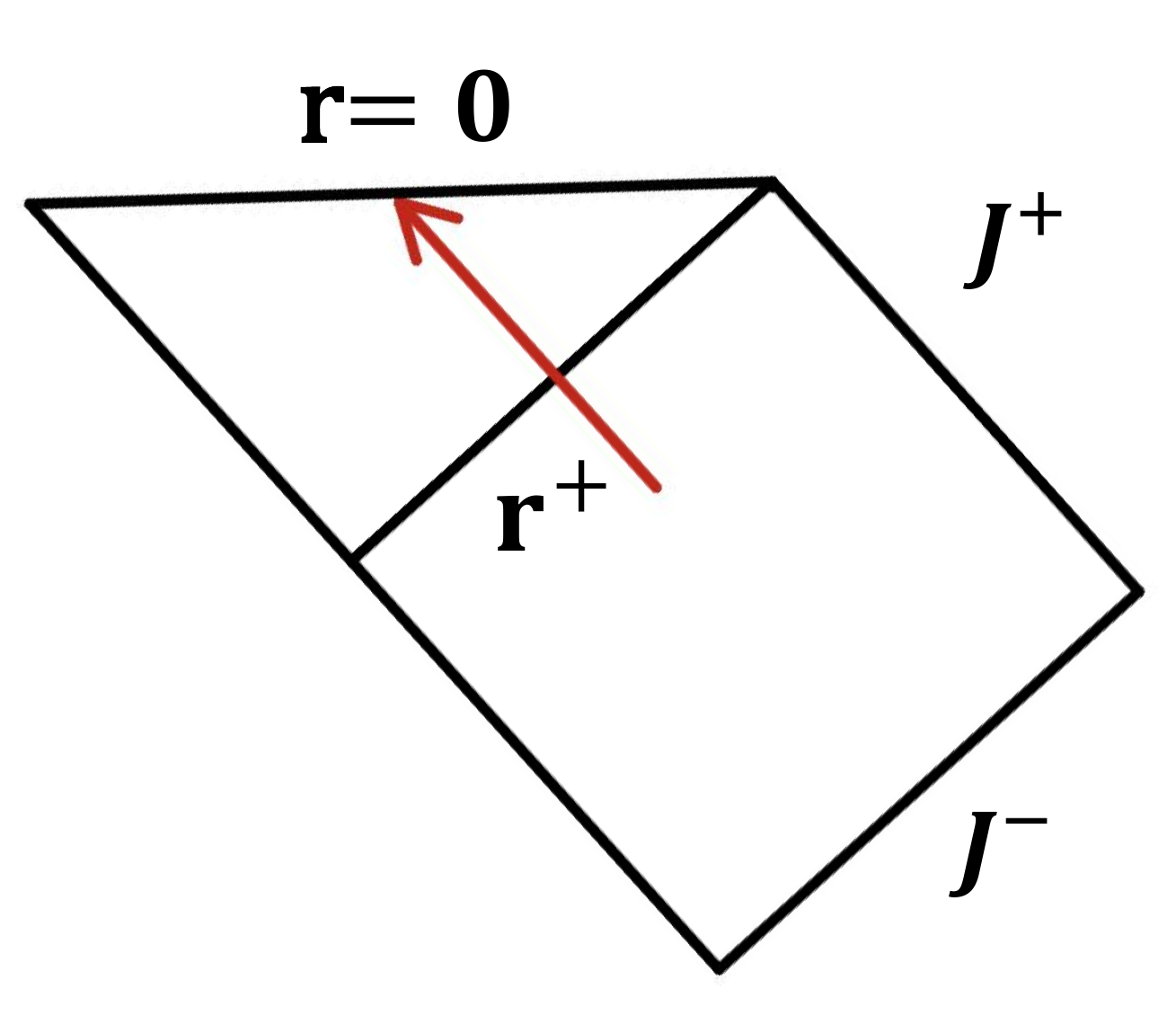}
\vskip -0.1truein
\caption{Penrose diagram of a static Schwarzschild black hole.}
\vskip -0.1truein
\label{SCPD1}
\end{figure}

\begin{figure}[htp]
\centering
\vskip -0.1truein
\includegraphics[width=0.35\textwidth]{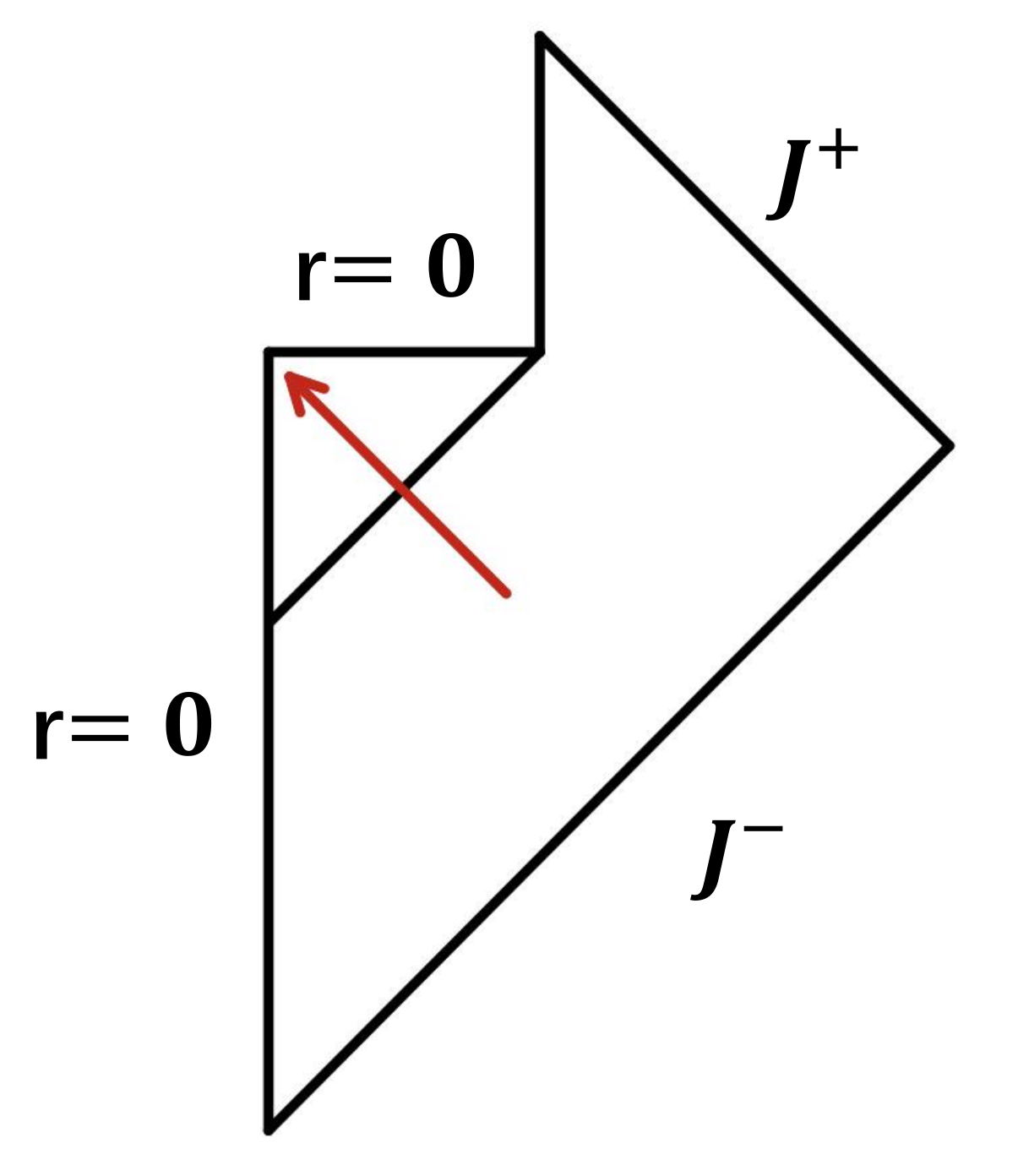}
\vskip -0.1truein
\caption{Penrose diagram of an evaporating Schwarzschild black hole.}
\vskip -0.1truein
\label{SCPD2}
\end{figure}

The information loss problem
may be easily described using Penrose diagrams. The Penrose diagram in Fig.~\ref{SCPD1} depicts a static Schwarzschild (non-rotating, uncharged) black hole.
In this case, the singularity is spacelike, and it is clear that information, from within the horizon, propagating along null or timelike geodesics cannot reach the outside Universe. 
The Penrose diagram in Figure \ref{SCPD2} describes an evaporating Schwarzschild black hole. The evaporating procedure of a Schwarzschild black hole does not change the fact that the horizon is along a null direction and hence does not improve the situation.

\begin{figure}[htp] 
\centering
{\includegraphics[width=0.30\textwidth]
{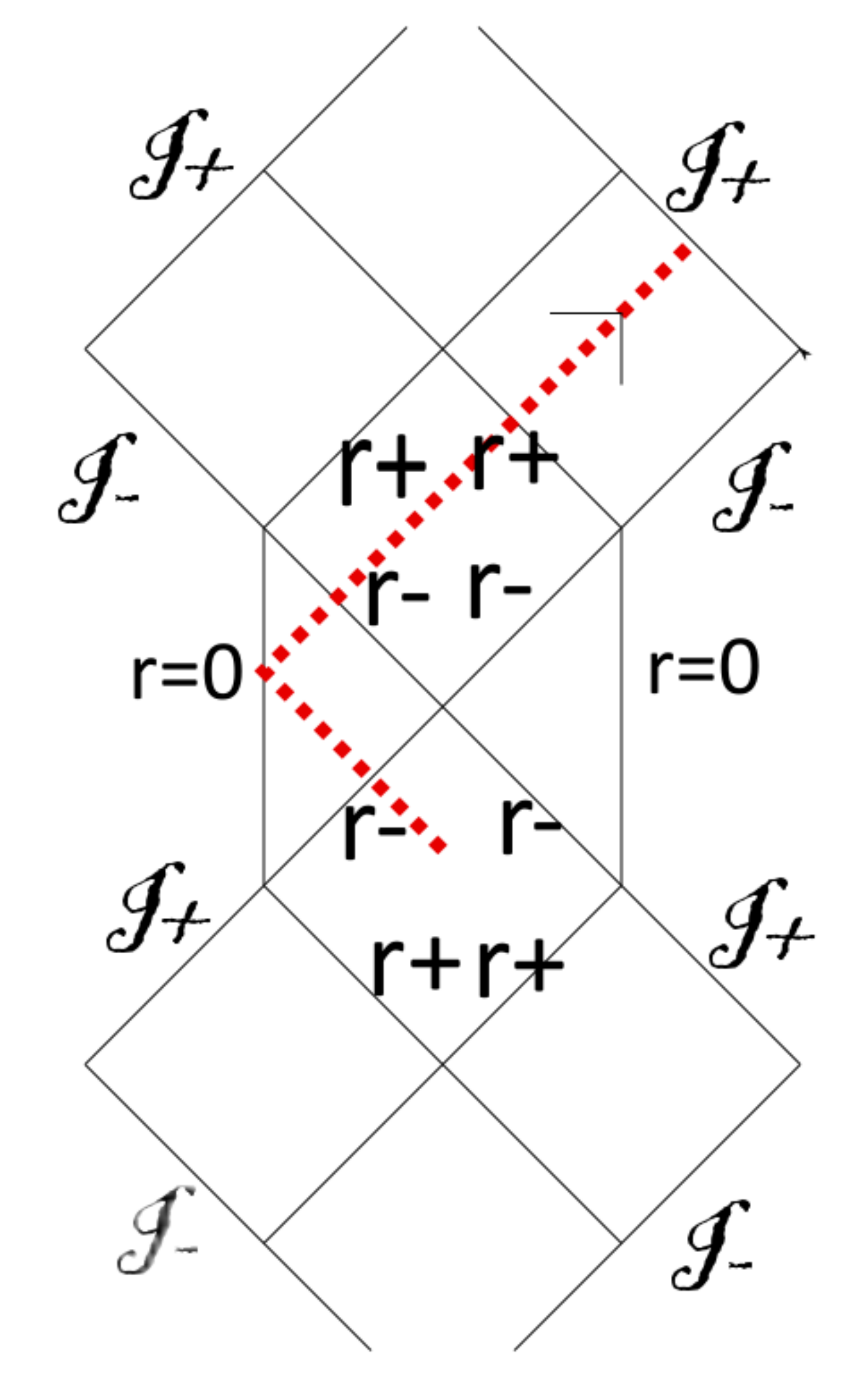}}
\vskip -0.1truein
\caption{Penrose diagram for stationary Kerr black hole.}
\label{Kerr}
\end{figure}

Similarly, Figure \ref{Kerr} is the 
Penrose diagram of a maximally extended 
rotating black hole whose singularity is 
timelike. Although information 
 propagating along null rays in such a Penrose diagram can exit a future horizon, it does so only by entering another Universe. Therefore, the information loss problem is not resolved in the present Universe. 

\begin{figure}[htp]
\centering
\includegraphics[width=0.45\textwidth]{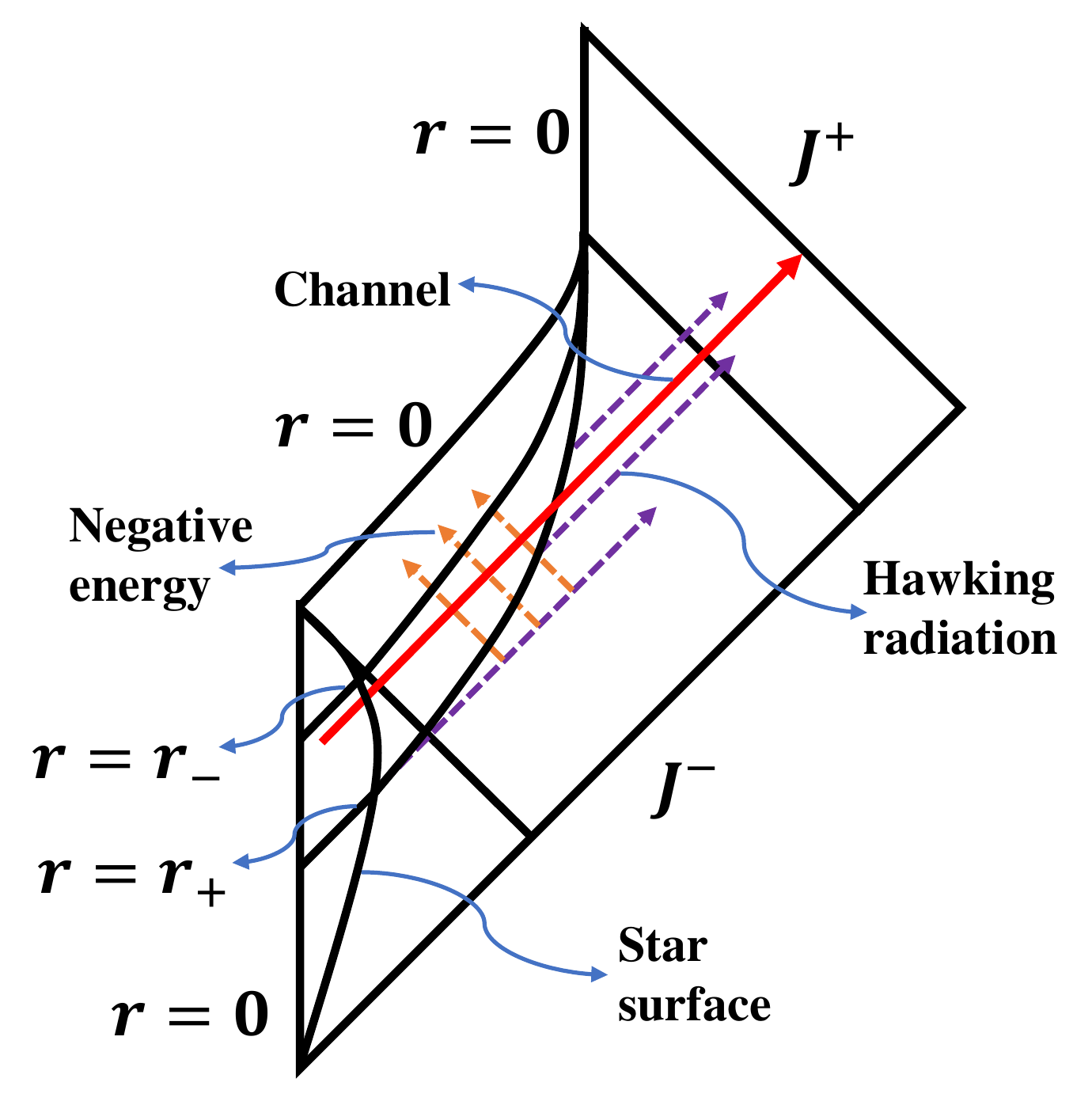}
\caption{Penrose diagram for an evaporating rotating (charged) black hole. This Penrose diagram describes a rotating (charged) star first collapsing into a rotating (charged) black hole. Then the black hole begins to evaporate, and both its inner and outer horizons at $\theta=0$ follow timelike trajectories. After the black hole completely evaporates away, the spacetime is connected with a flat spacetime.}
\label{Observers}
\end{figure}

Based on the results of Eqs.~(\ref{ds2}), (\ref{evaphorManu}), and (\ref{n2}), 
we can draw the Penrose diagram for the evaporating charged, rotating black hole (Fig.~\ref{Observers}). 
As one can see, there is a channel (outgoing null rays in red) which can carry classical as well as quantum information 
all the way from the 
interior to infinity, crossing the timelike horizon along the way.

We assume $\theta =0$ in Eq.~(\ref{metric2}) to simplify our calculations, it does not restrict our conclusion that there exists at least one such outgoing information channel. Even if future work shows that this channel is unique because its special azimuth, there is no limit in principle to the amount of information that can exit through this channel. Since our major calculations are carried out for a specific azimuth $\theta =0$, its generalization to black holes in higher-dimensional spacetimes should be straightforward.

\section{Discussion}

We have shown in this article that the apparent horizon and the region close to $r=0$ of an evaporating charged rotating black hole follow timelike trajectories. This implies that there exists a channel through which classical or quantum information can escape. Since proposals such as the fuzzball and firewall deal purely 
with quantum information, and in the context of Hawking radiation,
there is no contradiction between those
and our results here \cite{fuzz,Braunstein2009,Almheiri2013}. On the contrary, our approach complements the others. 

It is known that matter flowing into the dynamic charged and rotating black hole may cause instability due to mass-inflation, resulting in the nature of the singularity changing \cite{israel,rossi}. However, we do not anticipate such a problem in our case since (a) our results pertain to the region of low curvature (not singularity),
and (b) we have outgoing as opposed to infalling matter. That said a careful study of the effect for the spacetime under consideration is warranted and we hope to report on this in the future.

Since the horizon of a stationary black hole is null and the evaporating procedure is a quantum effect, emergence of such a classical channel is not a pure classical phenomenon. If the evaporation process stops for any reason before the black hole evaporates completely, our Fig.~\ref{Observers} would change and revert back to Fig.~\ref{Kerr}. However, since the Hawking temperature remains non-zero throughout, we do not consider this possibility. 

Although our work does not
completely resolve the problem of information loss, it complements
other approaches in opening up a new information recovery channel.
More work needs to be done to determine the extent of information that can be extracted via this channel, and its interplay with the 
quantum resolution of the singularity.

\appendix

\section{The metric in Parikh and Wilczek's paper represents a white hole}

To our knowledge, such an analysis was first done by
in \cite{Parikh1999} for non-rotating black holes. 
However, contrary to their assuming the metric that they used to be that of an evaporating black hole, it is in fact that of a charged white hole. They study the metric \cite{Parikh1999}
\begin{eqnarray}
ds^2 = -\biggl(1-\frac{2Mr-Q^2}{r^2}\biggr)du^2 - 2du\, dr + r^2 ( d\theta^2 + \sin^2\theta\, d\phi^2), 
\label{Parikh}
\end{eqnarray}
where $M=M(u)$ and $Q=Q(u)$ denote smooth decreasing functions of the retarded time $u$. It is easy to see that Eq.~(\ref{Parikh}) represents a white hole by taking $Q=0$ \cite{Poisson2004}. To show that Eq.~(\ref{Parikh}) is the metric of a white hole, we now calculate the expansion of the null normal vector of this metric.

Since this metric is spherically symmetric, it is easy to figure out that the outward null normal vector $l^\mu$ and the inward null normal vector $n^\mu$ for this metric are
\begin{equation}
l^\mu=(0,1,0,0)  \;\;\;\; \text{and} \;\;\;\; n^\mu= \Bigl( 1,-\frac{r^2 -2Mr+Q^2}{2 r^2},0,0 \Bigr).
\end{equation} 
It is easy to check that
\begin{equation}
l^\mu l_\mu =0 , \;\;\; n^\mu n_\mu=0 \;\;\; \text{and} \;\;\; l^\mu n_\mu=-1 .
\end{equation}
Finally, we calculate the expansion for both $l^\mu$ and $n^\mu$ by $\theta^{(A)} = A_{\mu ; \nu} \sigma^{\mu\nu}$, where $\sigma^{\mu\nu} = g^{\mu\nu} + l^\mu n^\nu + n^\mu l^\nu$, to obtain
\begin{equation}
\theta^{(l)} = \frac{2}{r} \;\;\;\; \text{and}   \;\;\;\;  \theta^{(n)} = - \frac{ r^2 -2Mr+Q^2 }{r^3} .
\label{expansionP}
\end{equation}

From Eq.~(\ref{expansionP}) we know that $\theta^{(l)}>0$ and $\theta^{(n)}<0$ when $r$ is larger than the apparent horizon. However, it is the vanishing of $\theta^{(n)}$ that defines an apparent horizon. Thus, the Vaidya-type metric in Parikh and Wilczek's paper represents a white hole.

\section{The metric in our manuscript represents a black hole}

To show that our metric represents an evaporating black hole, we calculate the expansion of the null normal vectors for the rotating charged Vaidya-type black hole, see Eq.~(1) in the manuscript.

Suggested by the rotating charged stationary black hole \cite{Newman}, we may conjecture that the outward null vector $l^\mu$ and the inward null vector $n^\mu$ for the rotating charged Vaidya-type black hole are
\begin{equation}
l^\mu=\Bigl( \frac{r^2 +a^2}{\sigma^2},\frac{\triangle}{2 \sigma^2},0,\frac{a}{\sigma^2} \Bigr) \;\;\;\; \text{and} \;\;\;\; n^\mu=(0,-1,0,0) .
\end{equation} 
where $\sigma^2=r^2 +a^2 \cos^2 \theta$ and $\triangle =r^2 +a^2-2Mr+Q^2$.
It is easy to check that

\begin{equation}
l^\mu l_\mu =0 , \;\;\; n^\mu n_\mu=0 \;\;\; \text{and} \;\;\; l^\mu n_\mu=-1
\end{equation}
for the Vaidya-type metric in our manuscript. Then we calculate the expansion for both $l^\mu$ and $n^\mu$ by $\theta^{(A)} = A_{\mu ; \nu} \sigma^{\mu\nu}$, where $\sigma^{\mu\nu} = g^{\mu\nu} + l^\mu n^\nu + n^\mu l^\nu$. Finally, we obtain

\begin{eqnarray}
\theta^{(l)} = \frac{ r \triangle + 2 \, a \,a'(u) (r^2 + a^2 \cos ^2(\theta ))}{(r^2 + a^2 \cos ^2(\theta ))^2}  \;\;\;\; \text{and} \;\;\;\;
\theta^{(n)} = -\frac{2 r}{r^2 + a^2 \cos ^2(\theta )} .
\label{expansionAp}
\end{eqnarray}

From Eq.~(\ref{expansionAp}) we see that $\theta^{(l)}>0$ and $\theta^{(n)}<0$ when $r$ is larger than the apparent horizon and $\theta^{(l)}=0$ defines the apparent horizon. Thus, the Vaidya-type metric in our manuscript represents a black hole.

The apparent horizon for a stationary rotating charged black hole is defined by $\triangle=0$. Here, our calculations show that the time dependence of the angular momentum $a(u)$ makes the apparent horizon shift slightly from the stationary case with the size of the shift depending on the rate of change of $a(u)$ with respect to $u$.

If we insert our linear functions for $M$, $Q$ and $a$, Eqs.~(3-6) in the manuscript, into the expansion of the outward null normal vector and suppose the rate of change is very slow ($\mu$ is very small), we can obtain the corrected apparent horizon for $\theta=0$ as

\begin{equation}
R_\pm = r_\pm + \frac{a \lambda_1 (r_\pm^2+a^2)}{r_\pm^2-(M+2a\mu \lambda_1)r_\pm} \mu + O(\mu^2) .
\end{equation}
Since we have proved $r_\pm$ is timelike, the apparent horizon of a slowly evaporating rotating charged black hole should also be timelike.

\end{document}